\documentclass[twocolumn]{aastex631}

\newcommand{\ie}{\emph{i.e.},~}
\newcommand{\eg}{\emph{e.g.},~}

\usepackage{amsmath}
\shorttitle{How the Cookie (Gas-rich Disk) Crumbles}
\shortauthors{Orr \& Rennehan}

\begin{document}

\title{How the Cookie Crumbles: A Model for Star-forming Clumps in High-redshift Disk Galaxies}

\author[0000-0003-1053-3081]{Matthew E. Orr}
\affiliation{Center for Computational Astrophysics, Flatiron Institute, 162 Fifth Avenue, New York, NY 10010, USA}
\affiliation{Department of Physics and Astronomy, Rutgers University, 136 Frelinghuysen Road, Piscataway, NJ 08854, USA}

\author[0000-0002-1619-8555]{Douglas Rennehan}
\affiliation{Center for Computational Astrophysics, Flatiron Institute, 162 Fifth Avenue, New York, NY 10010, USA}



\begin{abstract}

We present a simple model for the number distribution of maximally star-forming clumps in rotating disk galaxies, at high-$z$ with high gas surface densities.  By combining assumptions surrounding marginal stability of disks against gravitational fragmentation and collapse (\ie Toomre's $Q\approx 1$), star cluster formation efficiency scaling with local gas surface density, and star formation rates being tied to the relevant local dynamical/free-fall times, we find a star-forming clump distribution of $N_c(> \dot M_\star) \propto \dot M_\star^{-4/3}$ when assuming a power-law form for the gas surface density profile, and a numerically integrable relation for arbitrary gas disk profiles.  We compare this model with recent high-redshift observations of lensed clumpy star-forming rotation-dominated galaxies, and find good agreement with the distribution of clump star formation rates and number of clumps.  Moreover, we argue that any rotation-supported galaxy should have a significantly higher number of identifiable star-forming clumps relative to dispersion supported objects at a similar mass as $N_c \sim (V_c/\sigma)^2$.

\end{abstract}



\section{Introduction} \label{sec:intro}

Disk galaxies dominate the star formation rate density of the local volume \citep[][]{Behroozi2013}, and generally do so in a mode of feedback-regulated steady state star formation \citep{Schinnerer2024}.  These galaxies exhibit a regime of smooth gas accretion in the galactic outskirts, which slowly migrates radially inward replenishing gas as it is depleted by star formation \citep{Rathaus2016}.  The timescales for gas accretion, radial gas migration, stellar feedback, and the orbital time are ordered hierarchically such that equilibria naturally arise, and departures from steady state are usually dampened over a few dynamical times \citep{Faucher-Giguere2018}.

In broad strokes, star formation regulates itself in disk galaxies at late times by locally balancing the rate at which turbulent energy is lost in gas with the injection of energy and momentum from stellar feedback \citep[see, \eg][]{Ostriker2011, Faucher-Giguere2013, Ostriker2022}.  Principally, for regulation on large scales, this is in the form of supernova feedback. Turbulent energy yielded from the inward radial migration of gas also plays a role in offsetting decay terms, but is generally subdominant to feedback \citep[][]{Krumholz2018,Forbes2023}.  There is an attractor state of a hydrostatically supported disk, whose scale height locally is approximately the driving scale of supernova turbulence \citep{Orr2022a}.  However, as the timescale for collapse/star formation and the delay between the formation of young stars and supernovae occurring are non-zero, there are necessarily local departures from equilibrium and cycles of star formation/feedback arise \citep{Benincasa2016, Orr2019, Furlanetto2022}.  By averaging over many star-forming regions (\eg azimuthally, or across entire disks), we recover our predicted expectation values for star formation rates in disks, often in the form of scaling relations like Kennicutt-Schmidt \citep{Schruba2010, Kruijssen2014}.

Disk galaxies are not a phenomenon restricted only to late times, with rotation-dominated galaxies seen possibly as early as $z \approx 9.1$\footnote{$\sim$540 Myr after the Big Bang} \citep{Tokuoka2022}, with even earlier hints of detections as early as $z \approx 10.6$ \citep[][]{Xu2024}.  Compared to disks in the local universe, these high-redshift objects are significantly more gas rich with $f_{\rm gas} \sim 0.5-0.9$, compared to $\sim 0.1$ for local systems, and are dynamically much hotter with $\sigma/V_c \sim 0.5-1$, compared with the much colder $\sim 0.05-0.1$ of nearby, gas-poor spirals \citep[see a recent review of the evolution of the properties of the star-forming ISM by][]{Tacconi2020}.  Importantly, these highly gas-rich, highly turbulent, but still rotation dominated, disks are among the largest, and most highly star-forming galaxies at these redshifts.

Regardless of redshift context, rotating gas disks in galaxies are well-characterized by dynamically arranging themselves to be marginally stable against gravitational fragmentation and collapse \citep[][]{Leroy2008, Forbes2014, Villanueva2021, Aditya2023}.  After all, structures that are highly unstable do not survive long: one tends to find only the remnants of such disks in the form of massive star-forming complexes or constellations of young stellar clusters.  Thus, the scales on which we would expect fragmentation/collapse to occur, when disks go unstable, is that of the critical Toomre-length \citep{Toomre1964}.  To an order-unity pre-factor, the critical length is the local scale height of the gas. And so, the clumpiness of disks, in terms of collapsing star-forming structures, is then related to the aspect ratio of the disk (\ie how many scale heights fit across it), $H/R \approx \sigma/\Omega R\sim \sigma /V_c$, following from $\Omega \equiv V_c/R$ and that the local scale height can be derived from the ratio of out-of-plane to in-plane (rotational) velocities, \ie $H = \sigma/\Omega$.  We immediately intuit then that dynamically hotter disks, \ie those with higher $\sigma/V_c$, ought to be `less clumpy', as they host fewer independent Toomre clumps.

Given that disk galaxies at high redshift are dynamically hotter, these systems can be modeled as a (small) collection of Toomre-clumps \citep{Behrendt2019, Lenkic2021, Renaud2021, Mandelker2024}. To build an intuition regarding the dynamics and evolution of high-$z$ disk galaxies, it is crucial then to understand the number of star-forming clumps and their star formation rates, as they significantly affect the overall structure and behavior of galactic disks at early times. This understanding also provides a framework for comparing the properties of high-redshift galaxies with those observed in the nearby universe by surveys like, \eg MaNGA or PHANGS.

In this Letter, we will develop a simple model for clumpy maximally star-forming disks, and apply it to the conditions of high redshift galaxies.  We specifically compare with the observed clumpy ``cosmic grapes'' disk galaxy found at $z \approx 6$ by \citet{Fujimoto2024}.  And broadly discuss the implications of the natural star-forming modes of gas-rich Toomre-clump disks, in the context of our model and zoom-in simulations of high-redshift galaxies, particularly the distribution function of star forming clumps/structures.

\section{Star-forming Clumpy Disk Model} \label{sec:model}

Developing a minimal model for a star-forming disk galaxy composed of collapsing Toomre-patches, we require three main components: the radial gas distribution and orbital structure, the distribution of Toomre-patches, and the star formation rate of those patches.

We generally parameterize the gas disk of our Toomre-patch galaxy as an exponential gas profile, \ie $\Sigma_g(r) = \Sigma_{g,0} e^{-r/R_g}$, where $\Sigma_{g,0}$ is the disk's central gas surface density and $R_g$ is the gas scale length (normalized such that the total gas mass $M_g = \int^\infty_0 \Sigma_g(r) 2 \pi r dr = 2 \pi \Sigma_{g,0} R_g^2$). Capturing the shape of the inner halo (gas + stars + dark matter), we parameterize the rotation curve of the disk as
\begin{equation}\label{eq:vc}
    V_c(r) = \begin{cases}
    V_c \left(\frac{r}{R_{\rm flat}}\right)^\alpha &  r < R_{\rm flat} \\
    V_c &  r > R_{\rm flat} \; , 
    \end{cases}
\end{equation}
where $V_c$ is the maximum/constant value of the rotation curve, $R_{\rm flat}$ is the radius at which the maximal value occurs, and $\alpha$ parameterizes the shape of the rising portion of the rotation curve.

We also assume that the entire disk has a constant value of Toomre-Q, \ie 
\begin{equation}\label{eq:Q}
    Q \approx Q_0 = \frac{\kappa(r) \sigma f_g}{\pi G \Sigma_g(r)}
\end{equation}
where $\sigma$ is the local turbulent velocity dispersion (or local sound speed, if the gas is primarily thermally supported), $f_g$ is the local gas fraction (to account for the contribution of the self-gravity of stars/dark matter within the gas patch to its instantaneous stability, or lack thereof, against fragmentation/collapse), and $\kappa^2 = \frac{2 \Omega}{r} \frac{d}{dr}(r^2 \Omega)$ is the epicyclic frequency ($\Omega = V_c(r)/r$ being the orbital frequency). And for this parameterization of rotation curve, $\kappa$ becomes
\begin{equation}
    \kappa^2 = \begin{cases}
        2 (\alpha + 1) \Omega^2 & r < R_{\rm flat} \\
        2 \Omega^2 & r > R_{\rm flat} \; .
    \end{cases}
\end{equation}
For extremely gas-rich galaxies, where $f_{\rm gas} \approx 1$, we argue that adopting more detailed stability parameters (\eg those of \citealt{Rafikov2001} or \citealt{Romeo2013}) for our simple model is unnecessary given that these are essentially one-fluid systems.

At a most basic level we model the star formation rate of a `clump' of gas of mass $M_c$ in the ISM as occurring with some efficiency $\epsilon$ per free-fall time $t_{\rm ff}$ as
\begin{equation}\label{eq:sfrdef}
    \dot M_\star = \epsilon(\Sigma_g) M_c/t_{\rm ff} \; ,
\end{equation}
where we take $t_{\rm ff} \equiv \sqrt{3\pi/32 G \rho}$, with $\rho$ as the mid-plane density $\rho = \Sigma_g/(Hf_g)$ and have assumed a model where the star formation efficiency follows that of \citet{Grudic2018}. We connect the initial clump mass (before the fragmentation/collapse that results in the star formation episode) with the local gas surface density assuming that the maximal isotropic instability scale is the Toomre-patch of size the local scale height in the disk.  Assuming this, we have $M_c = \pi H^2 \Sigma_g(r)$.  And for a (weakly self-gravitating) patch of gas with a vertical velocity dispersion $\sigma$, orbiting with circular velocity $V_c$, the local scale height is $H = \sigma r/V_c (= \sigma/\Omega)$. 

In high-resolution simulations of star-forming clouds with a wide range of gas surface densities, \citet{Grudic2018} found that the integrated star cluster formation efficiency (\ie the final mass of stars formed per initial gas mass of the cloud) scaled only with local (cloud) gas surface density as
\begin{equation}
    \epsilon(\Sigma_g) = \left( \frac{1}{\epsilon_0} + \frac{\Sigma_{\rm crit}}{\Sigma_g} \right)^{-1} \; ,
\end{equation}
where $\epsilon_0$ is the saturation efficiency (\ie when $\Sigma_g \gg \Sigma_{\rm crit}$; $\epsilon_0= 0.77$) and $\Sigma_{\rm crit}$ is the critical surface density for star cluster formation efficiency to saturate (taken to be $\Sigma_{\rm crit} = 2800$ M$_\odot$ pc$^{-2}$).

Combining the equation for the free-fall time, making use of our assumption for a constant Toomre-Q (Eq.~\ref{eq:Q}), the parameterized form of the rotation curve (Eq.~\ref{eq:vc}), and the equation for the local scale height into Eq.~\ref{eq:sfrdef}, we find a form for the star formation rate $\dot M_\star$ of a Toomre-patch in our model of
\begin{equation} \label{eq:sfr} \begin{aligned}
    \dot M_\star(r) = \epsilon(\Sigma_g) \frac{4\pi^2}{18^{1/4}} Q_0^{3/2} \left( \frac{G}{f_g} \right)^2 \left( \frac{r \Sigma_g}{V_c} \right)^3 \\ \; \times \begin{cases}
        (\alpha+1)^{-3/4} (r/R_{\rm flat})^{-3\alpha} & r < R_{\rm flat} \\
        1 & r > R_{\rm flat}
    \end{cases} \;, 
\end{aligned}
\end{equation}
which is expressed purely in terms of the radially dependent functions $\epsilon(\Sigma_g(r))$ and $\Sigma_g(r)$.

\subsection{Clump Counting}

\begin{figure}[ht!]
\plotone{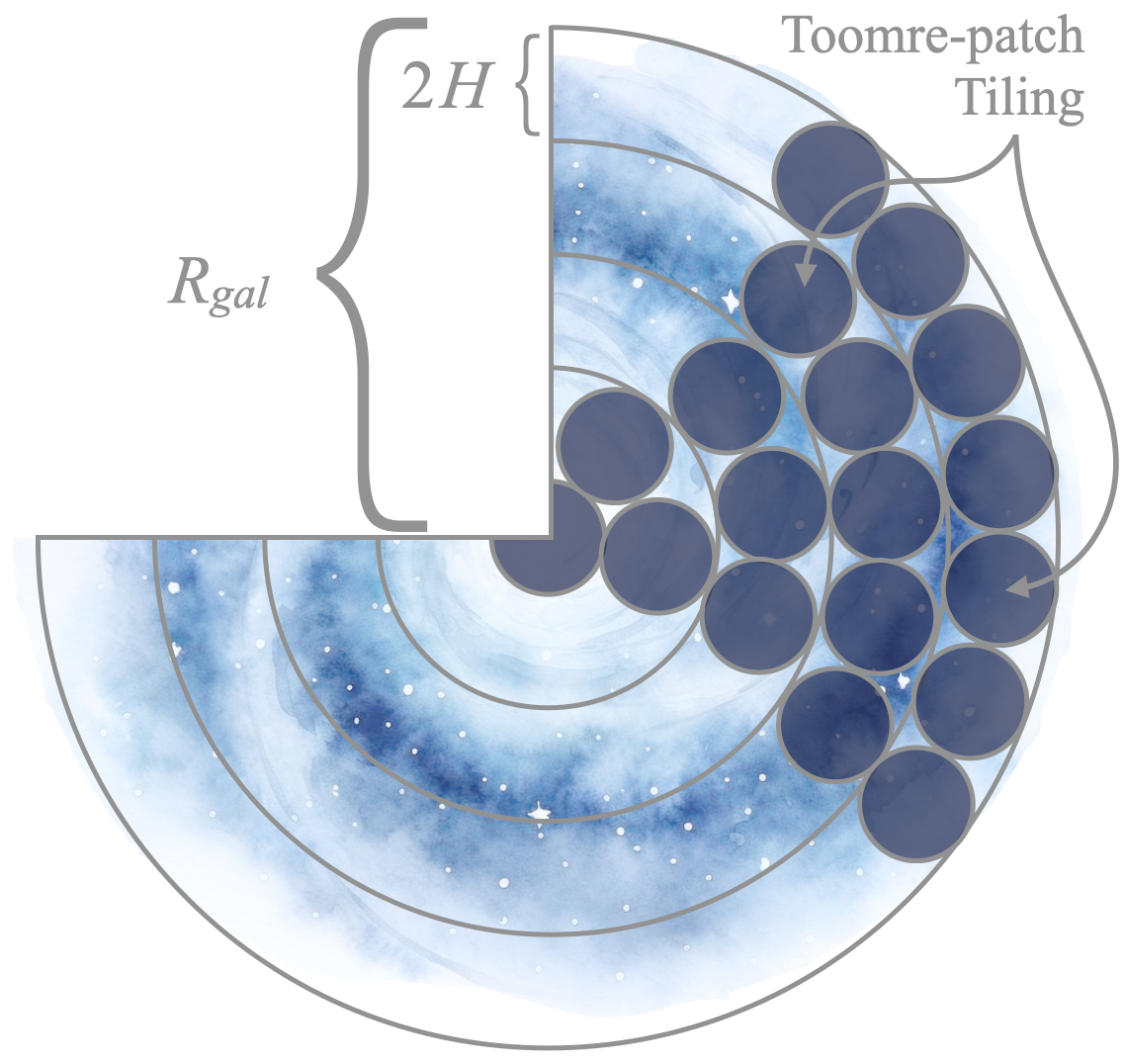}
\caption{Cartoon illustrating our simple clump tiling for counting the number of clumps in a disk. Starting from the outer edge $R_{gal}$, we tile the annulus with maximal Toomre-patches of radius $H$ and then step in by $\Delta r = 2 H$ (which iteself changes with radius) until reaching the center.  We calculate the star formation rate at the center of each annulus, assigning it to all clumps at that galactocentric radius.
\label{fig:cartoon}}
\end{figure}

Similarly to the star formation rate, for our simple model we take the size of clumps to be purely radially dependent, scaling as the local gas scale height.  And the number of clumps, \ie $N_c(r)$, too, as filling a ring of width $2H$ at galactocentric radius $r$,
\begin{equation}\label{eq:nclump} \begin{aligned}
    N_c(r) &\approx \frac{4 \pi r H}{\pi H^2} = \frac{4 r}{H}  \\& = \frac{4\sqrt{2} f_g V_c^2}{\pi Q_0 G r \Sigma_g}\begin{cases}
        \sqrt{\alpha +1} (r/R_{\rm flat})^{2\alpha} & r < R_{\rm flat} \\
        1 & r > R_{\rm flat}
    \end{cases}
    \end{aligned}
\end{equation}
using our assumption that Toomre's $Q$ is a constant $Q_0$.

It is not possible to analytically compute the number of clumps exclusively as a function of star formation rate when assuming an exponential profile for the gas disk.  However, loosening that form to that of a power law, \ie $\Sigma_g(r) = \Sigma_{g,0} (r/R_g)^{-\beta}$, allows us to do so.  To compare with observations of the cumulative distribution of star-forming clumps, we aim to calculate the number of clumps with star formation rates greater than a particular $\dot M_\star$, \ie 
\begin{equation}
    N_c(> \dot M_\star) = \int^\infty_{\dot M_\star} \frac{d N_c(\dot M_\star)}{d \dot M_\star} d\dot M_\star \; .
\end{equation}

We can express the star formation rate in purely radial terms, with $\dot M_\star \propto r^{3-3(\alpha+\beta)}$ and thus $d \dot M_\star/dr \propto r^{2-3(\alpha+\beta)}$, from assuming a power law form for $\Sigma_g(r)$, and a saturated/constant star formation efficiency $\epsilon(\Sigma_g) \rightarrow \epsilon_0$ in the $r<R_{\rm flat}$ case of Eq.~\ref{eq:sfr}.  Allowing us to substitute into 
\begin{equation}
    N_c(> \dot M_\star) = N_c(< R(\dot M_\star)) = \int^R_0 \frac{2 \pi r dr}{\pi H^2}\; ,
\end{equation}
and following the appropriate change of variables and integration, in terms of $\dot M_\star$, we find a closed-form solution of
\begin{equation}  \begin{aligned}
    N_c(> \dot M_\star) = \frac{2(\alpha +1) f_g^2 V_c^4}{(2\alpha + \beta -1) \pi^2 Q_0^2 G^2 \Sigma_{g,0}^2 R_{\rm flat}^{4\alpha}R_g^{2\beta}} \\
    \times \left[ \frac{4 \pi^2 \epsilon_0 Q_0^{3/2} G^2}{18^{1/4}f_g^2} \left(\frac{\Sigma_{g,0}R_g^\beta R_{\rm flat}^\alpha}{(\alpha+1)^{1/4} V_c}\right)^{3} \right]^\frac{2(2\alpha+\beta-1)}{3(\alpha+\beta-1)} \\
    \times \dot M_\star^\frac{-2(2\alpha+\beta-1)}{3(\alpha+\beta-1)}\; .
    \end{aligned}
\end{equation}
This solution for the number of clumps simplifies dramatically when taking $\alpha = \beta = 1$ to
\begin{equation} \label{eq:analyticform}
    N_c(> \dot M_\star) = \frac{2^{8/3}}{18^{1/3}} \left(\frac{\pi \epsilon_0 G}{f_g}\right)^{2/3} (\Sigma_g^0 R_g)^2  \dot M_\star^{-4/3}\; .
\end{equation}

So as to compare with observations, we also calculate the number of clumps at a given radius and their star formation rate from equations \ref{eq:sfr} \& \ref{eq:nclump}, assuming an exponential profile for the gas. To do so, we start from the outer edge of the disk, and tile inwards in steps of $\Delta r = 2H$.  Figure~\ref{fig:cartoon} shows a cartoon of the general scheme of Toomre-patch tiling for our calculation. We calculate the star formation rate given the clump properties at each step in radius.

When comparing the analytic solution (Eq.~\ref{eq:analyticform}) assuming a power-law gas surface density profile with the numerically integrated exponential gas disk profile, one must include a $(1-2/e)$ term to the gas surface density, such that $\Sigma_g(r) = (1-2/e) \Sigma_{g,0} R_g/r$, to normalize the gas mass within $R_g$ to the same as that of the exponential profile.


\section{Comparison with Gas-rich High-z Galaxy Conditions} \label{sec:res}
\begin{figure}
    \centering
    \includegraphics[width=\linewidth]{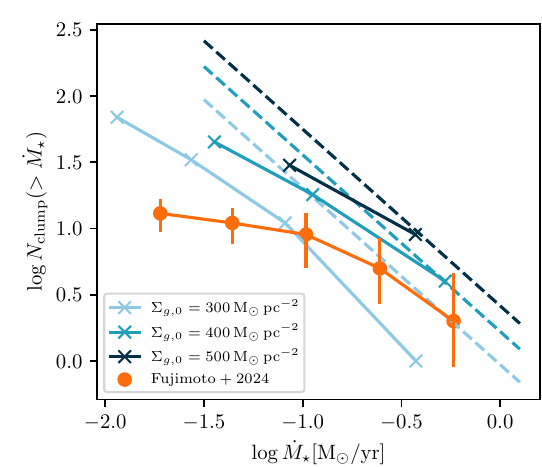}
    \caption{Cumulative star-forming clump distribution $N_c(>\dot M_\star)$, comparing our clumpy disk model with observations of a $z \approx 6$ star-forming disk by \citet{Fujimoto2024}. Solid blue lines indicate numerically integrated results of our model (Eqs.~\ref{eq:sfr} \& \ref{eq:nclump}) assuming an exponential gas disk profile with the following parameters, matched to observations: $R_{gal} = R_{gas} =R_{flat} = 0.7$ kpc, $\alpha=1$, $\Sigma_{g,0}=3-5\times 10^2$ M/pc$^2$, $V_c = 70$ km/s, $f_g=1$, $Q_0 = 1$. Dashed lines indicate analytic solution of our model, assuming a power-law gas surface density profile (Eq.~\ref{eq:analyticform}), assuming the same parameters, except setting $\epsilon_0 = 0.1$. A clumpy star-forming (exponential) disk model generally exhibits a similar star formation distribution to the high-$z$ observations.}
    \label{fig:grapescompare}
\end{figure}


We make direct comparison between our simple clumpy disk model and observations of the ``cosmic grapes'' by \citet{Fujimoto2024}, a highly lensed (to a spatial resolution of 10-60 pc) $z\approx 6$ \textit{very clumpy} star-forming disk galaxy.  Their observations provide all the parameters needed for our clump-counting: a physical scale, a rotation curve, and an estimate for the gas surface density profile (and $f_{\rm gas}$). 

Figure~\ref{fig:grapescompare} shows the result of inputting the ``cosmic grapes'' parameters from \citet{Fujimoto2024} into Eqs.~\ref{eq:sfr} and \ref{eq:nclump} and numerically integrating the clump distribution as a function of SFR: $R_{gal} = R_{gas} = R_{flat} = 0.7$ kpc, $\alpha=1$, $\Sigma_{g,0}=4 \pm 1\times 10^2$ M/pc$^2$, $V_c (V_{\rm flat}) = 70$ km/s, and $f_g=1$. Further, we take $Q_0 = 1$.  Adopting these fiducial parameters, our general model describes the clump star formation rate distribution found in the observations, both in normalization and in the slope.  We also plot the solution in Eq.~\ref{eq:analyticform}, choosing $f_{\rm gas}=1$ and $\epsilon_0=0.2$.  Otherwise adopting the same parameters as the numerically integrated Eq.~\ref{eq:sfr} and \ref{eq:nclump}, and including the aforementioned $\Sigma_{g,0} \rightarrow (1-2/e)\Sigma_{\rm g,0}$.  The gas profile does not fall as quickly in the outskirts as the exponential, so it's unsurprising that there is an overabundance of star-forming clumps at large radii (low SFRs).  However, it does give us an intuition for the slope the relation.

Moreover, the high-$\dot M_\star$-end slope of the numerically integrated model is very close to that yielded by the simplifying assumption of a power-law gas disk in the model, \ie $N_c(>\dot M_\star) \propto \dot M_\star^{-4/3}$ when $\alpha = \beta = 1$.  This suggests that this object is indeed dominated by efficiency saturated star cluster-forming regions.  As well, both our numerically integrated exponential profile and power-law profile assume a 100\% occupation fraction of observable star-forming clumps per Toomre patch at any given time, which likely breaks down towards the outskirts of galaxies, where asymmetric gas accretion/radial flows are important- explaining in part the growing discrepancy at lower SFRs.  And so, in our estimation, our model represents a maximally star-forming clumpy disk at these times.


\section{Discussion}


\subsection{Clumps and Rotation Support}

The largest coherent star-forming regions in any galaxy are set by the local scale height $H$.  This follows from the simple argument that if the crossing time (thermal or turbulent, whichever is shorter) of any structure is longer than the relevant timescales for star formation (\eg the free-fall time $t_{\rm ff}$), then that structure must contain independently collapsing/fragmenting substructures within which the crossing time is shorter than the relevant star formation timescales.  And these timescales are matched to order unity in gas-rich environments at the local scale height as $t_c = H/\sigma \approx r/V_c$, and $t_{\rm ff} = \sqrt{3\pi/32 G \rho} \approx 0.81 \sqrt{Q_0/f_g\sqrt{1+\alpha}} \times r/V_c$ when we take $\rho = \Sigma_g/H$ and use our definition of Toomre-Q (Eq.~\ref{eq:Q}, assuming it to be a constant $Q_0$).

And so, to zeroth-order, the aspect ratio of a galaxy, \ie $H/R_{\rm gal}$, immediately informs us as to the \textit{rough} number of star-forming clumps to expect in a gas-rich galaxy: namely, \textit{a few} or \textit{many}.  In dispersion-supported galaxies $\sigma/V_c \sim 1$ suggesting the typical intuition that the typical scale of star-forming HII regions is $H \sim \sigma r / V_c \sim r$, \ie the whole galaxy is $\sim$one star-forming region. Whereas for rotationally supported disks which are dynamically cold, $\sigma/V_c \ll 1$, and so many scale lengths fit across the disk ($r/H \sim r/(\sigma r / V_c)\sim V_c/\sigma \gg 1$), \ie the galaxy contains \textit{many} independently evolving star-forming regions.

Seeing the importance of developing this intuition, \citet{Fujimoto2024} makes a comparison between their ``cosmic grapes'' at $z \approx 6$ and similarly massed cosmological zoom-in simulations from the FIRE-2 suite \citep{Wetzel2022} at that redshift.  They found that their observed rotating gas-rich disk galaxy (with $\sigma/V_c \approx 1/2.94$) seemed to have a high number of star-forming clumps compared to the FIRE-2 simulations, perhaps anomalously so.  Critically, however, the FIRE galaxies at this redshift, though similarly massed, are not rotationally supported.  Following our previous line of argument, we should strongly expect that dispersion supported galaxies (theoretical or not) should host many fewer independent star-forming regions (``clumps'') at the same level of star formation. In some sense, it is simply morphologically impossible for the FIRE galaxies to match the expected clump number of the ``cosmic grapes''\footnote{Really, there is a selection effect in the FIRE-2 sample to late-forming systems that can affect our understanding of the timing of galaxy evolution processes (see e.g. \citealt{Rennehan2024}).}


\subsection{Are these ``Feedback Free'' Disks?}

With the avalanche of high-$z$ starburst observations from JWST, there has been a renewed focus on high gas density, high star formation efficiency environments \citep{Cole2023, Andalman2024, Robertson2024, Marques-Chaves2024, Menon2024}. One such framework is the \citet{Dekel2023} ``Feedback-free starburst'' (or FFB galaxies) at extremely high redshift \citep[see also][]{Li2024}.  In these environments, extremely low metallicities result in weak (relatively speaking) radiation pressure coupling and stellar winds from massive stars, and the collapse/star formation timescales are so short as to entirely preclude supernovae as a meaningful feedback channel to regulate the starbursts.

For constant-$Q$ disk with a linearly rising rotation curve, the free-fall time ends up being constant across the disk (recalling $t_{\rm ff} \propto r/V_c$). In the cosmic grapes, the free-fall time would then be $\sim$9.5 Myr  everywhere.  This is longer than the timescale for core-collapse supernova feedback to begin, and so the high star formation efficiencies are related to the high gas surface densities, but these systems are not exactly ``feedback-free'', and are expected to behave more in-line with local extreme starbursts (that do have high $\sim$few $\times 10^2$ M$_\odot$ pc$^{-2}$ gas surface densities; \citealt{Lenkic2024}).

\subsection{Feedback and Wind-driving from High-Gas Fraction Disks}

Given the extreme star formation in these conditions, supernova-driven outflows should be highly prevalent, and as result the turbulence within the ISM driving somewhat \textit{low}, in relative terms.  Adopting the superbubble feedback model from \citet{Orr2022a}, where clustered supernovae drive superbubbles that expand to the disk scale height (driving ISM turbulence) before breaking out of the galaxy (driving outflows), we calculate the `effective' strength of supernova feedback that we would expect in the galaxy.  Using their fiducial parameters for superbubble feedback, and $f_g \approx 2/3$ and $\Omega \approx 90$ Gyr$^{-1}$ for the ``cosmic grapes'', in Eq.~20 of \citet{Orr2022a} we find $(P/m_\star)/(P/m_\star)_0 \approx 0.231$.  That is, the fraction of momentum per mass of young stars that would be expected to be injected as feedback in the ISM versus driving outflows is about 23\%, cleanly predicting that this ``powered breakout'' (PBO) results in ``weak'' supernova feedback in the ISM relative to ``strong'' outflows.  This again frames well with the picture that star formation is locally efficient in these systems given that feedback fails to effectively regulate star formation on $\sim$kpc scales.

\section{Summary \& Conclusion}

In response to recent JWST observations of a highly clumpy star-forming disk at $z\approx 6$, we pieced together a simple model for gas-rich star-forming disks comprised of packed Toomre-patches in order to understand the general distribution of maximally star-forming clumps in high redshift disks.
Our findings can be shortly summarized by the following points:

\begin{itemize}
    \item Our simple model of Toomre-patches, whose local star formation rates are primarily governed by their own internal free-fall times, in a marginally stable rotating disk do a remarkable job of reproducing the observed clump cumulative star formation rate distribution in the ``cosmic grapes'' found by \citet{Fujimoto2024}.


    \item The ``clumpy'' nature of observed high-gas fraction disks is to be expected given that the number of independent collapsing regions should scale as $N_c \sim (V_c/\sigma)^2$, essentially the inverse aspect ratio of the galaxy disk.  For the ``cosmic grapes'' the $V_c/\sigma \approx 3$ suggests a clump number on the order of $\sim$10, in-line with our order of magnitude scaling.  
\end{itemize}

Centrally, we find that these high-gas fraction, high-$z$ disks can be modelled as fairly straightforward systems of clumps collapsing from the largest disk instabilities containing the maximal amount of star-forming gas.  Feedback clearly is required to regulate these objects, but generally it is `prompt' feedback processes (\ie stellar winds, photoionization, radiation pressure, \textit{etc.}) that set the star formation efficiency of the clumps themselves given that supernovae do not have time to go off in these systems.  Only at later times, supernovae drive turbulence in and outflows from these near maximally efficient (in star formation) systems.  Follow-up spatially resolved observations of star-forming, high-$z$ disks will hopefully reinforce these findings, and the general picture that these disks are `simple' structures in these terms.

\begin{acknowledgments}

The Flatiron Institute is supported by the Simons Foundation.
This research has made use of NASA's Astrophysics Data System.

\end{acknowledgments}

\bibliography{library, cca_lib}{}
\bibliographystyle{aasjournal}



\end{document}